\documentclass[fleqn,10pt]{wlpeerj} 
\usepackage{todonotes}\reversemarginpar
\usepackage{color}
\usepackage[normalem]{ulem}
\usepackage{cleveref}

\newcommand{\add}[1]{\textcolor{red}{#1}}
\newcommand{\del}[1]{\textcolor{red}{\sout{#1}}}

\title{Predicting engagement in online social networks: Challenges and opportunities}

\author[1]{Farig Sadeque}
\author[1]{Steven Bethard}
\affil[1]{School of Information, University of Arizona, Tucson, AZ 85721}
\corrauthor[1]{Farig Sadeque}{farig@email.arizona.edu}

\begin{abstract}
Since the introduction of social media, user participation or engagement has received little research attention. In this survey article, we establish the notion of participation in social media and main challenges that researchers may face while exploring this phenomenon. We surveyed a handful of research articles that had been done in this area, and tried to extract, analyze and summarize the techniques performed by the researchers. We classified these works based on our task definitions, and explored the machine learning models that have been used for any kind of participation prediction. We also explored the vast amount of features that have been proven useful, and classified them into categories for better understanding and ease of re-implementation. We have found that the success of a technique mostly depends on the type of the network that has been researched on, and there is no universal machine learning algorithm or feature sets that works reasonably well in all types of social media. There is a lack of attempts in implementing state-of-the-art machine learning techniques like neural networks, and the possibility of transfer learning and domain adaptation has not been explored.  
\end{abstract}

\begin{document}

\flushbottom
\maketitle
\thispagestyle{empty}

\section*{Introduction}
Online social networks dominate the Internet, having established themselves as an integral part of human interaction and communication over the last decade. The term ``Online Social Network'' covers a wide range of services that provide online interaction -- from micro-blogging websites such as Twitter to support group based health forums such as DailyStrength. As these networks grew, much research has been done on them over the years. 
Yet, an intriguing aspect is mostly overlooked in such research: continued participation or engagement. Despite the fact that continued participation is one of the most significant contributors of social capital in online forums and social networks, there is little research on computational assessments of the level of engagement in popular social networking sites. Though engagement in large services like Twitter or Usenet newsgroups has been explored over the years, smaller networks, specifically support group or community based forums have received little to no attention.

Continued participation is a frequently researched topic across many industry sectors. A common way of talking about continued participation is in terms of \textit{churn} -- a portmanteau of \textit{change} and \textit{turn} -- which is the rate of loss of customers from a company's customer base to another company. Research on churn has a simple motivation: loss of customers is loss of revenue, and retaining a customer is much cheaper than winning a new one \citep{hadden2007computer}. Generally a company tries to identify a churning customer early in their lifecycle so that customer management departments can efficiently target these customers and provide incentives to prevent them from leaving the company. Among these industries, telecommunication sectors have contributed extensively in the research of churn among their customers \citep{Kim2004751,Gerpott2001249,Keaveney:1995,mozer1999churn,burez2009handling,dasgupta2008social}.

Consequently, continued participation is an important factor for social network services since they follow the same business model as the service providers in telecommunication sectors: you lose revenue when a customer leaves the network. However, in social networks, the threat is much more than monetary. As social networks thrive on the interactions among users, loss of users means loss of social capital within the service, which ultimately affects the sustainability of the service.
The strict definition of churn also typically does not apply to social networks, as users may or may not join another service after leaving the current one. Instead, terms like \textit{continued participation}, \textit{engagement}, \textit{attrition}, or \textit{defection} are more commonly used.
In the current article, we adopt the term \textit{engagement} since it encompasses the broadest range of phenomena.

Factors that influence engagement in social networks can vary from service to service. Graph based features can play a big role in predicting participation in those services which maintain an extensive architecture of relationships among the users like Facebook, whereas the frequency of activities plays a bigger role in the prediction task in services like forums and discussion boards. Demographic information, contents of texts, and timelines within user lifecycles can contribute significantly depending on the paradigm of the prediction task.  

In this work, we survey the aspects of social networks that have been shown by prior research to contribute to the prediction of engagement in different paradigms. We discuss the features and techniques that have already been used to predict future participation, along with the strengths and weaknesses of each approach. We also briefly discuss the challenges that are faced predicting future participation in social networks and possible research opportunities in this field.

\section*{Challenges in Predicting Engagement in Social Media}
Social network users invest their time, sharing views or opinions or simply participating in a discourse, without expecting any immediate return from the network \citep{constant1996kindness}. In sociology this type of activity is known as the ``Gift Economy'' \citep{ICT4DBibliography150}, which, in contrast to the the service or commodity economy, is not driven by exchanging service or commodities for monetary benefits, but rather is driven by the expectations of social contracts. Several motivations drive users to participate in this economy of gift transactions, for example, the expectations of future payback in terms of new information and social interaction, recognition as a source of valuable information from the peers or idea diffusion among other users in the community; and when these expectations are not met, users tend to leave the community, thus hurting the social capital of the network in the process. Social networking services lose revenue when users leave their network, just like other industries; but this loss of social capital poses a greater threat to the services as this threatens the survival of the social networks in the long run.

One of the challenges in predicting continued participation in online social networks is that there are no predefined ``triggering events'' \citep{gustafsson2005effects} in social networks as there are in telecom sectors. In telecommunication services, a subscriber is bound by a service contract or he buys credits before using the service. When the contract expires, or the credit dries up, churn is triggered based on the other factors like service quality, tariffs or poor customer experience. In social networks, users are weakly tied by a non-binding social contract \citep{constant1996kindness}. A user can leave a social network any time without incurring any kind of explicit monetary penalty, and can again join the network any time as there is low-entry barrier to join most social networks. This absence of triggering events makes it more difficult to predict continued participation in social networks than to predict churn in industries like telecom.

Another challenge while predicting continued participation in social networks is the diversity and the growth of the social networks \citep{Karnstedt:2011:EUF:2527031.2527051}. There are chatrooms, discussion boards, community forums, photo and video sharing websites, blogs, massively multiplayer online games, online courses and many others which accumulate two or more of these services into them. The inner structures of these services are highly diverse and complex. Discussion boards and blogs are mostly for sharing ideas and views by posts and replies in threads, and interpersonal communication among the users in these services are generally sparse, whereas chatrooms and online games depend mostly on the dense interpersonal communication among the users. Also, there are hierarchies of participation continuation in most of these services: a user can stop communicating with a single user or a set of users, or he can stop participating in a forum or a single thread, or he can leave the network entirely.

One other challenge that makes engagement prediction in social networks more difficult than predicting churn in the telecom sector is that in social networks, participation is a continuous process. A user does not suddenly drop off of a social network, it happens over a significant period of time. The amount of activity a user performs in a social network over a certain period of time is a good predictor of that user's future participation \citep{sadeque2015predicting,sadeque2016they}. There is no certain triggering event in social networks as there is in telecom services; a user may gradually decrease his or her participation in the community and eventually stop participating at all.   

Due to these challenges, engagement prediction in social networks is still largely unexplored, and thus represents a major research opportunity in this field. There have been a few works on predicting future participation in popular paradigms like micro-blogging (e.g., Twitter; \cite{DBLP:journals/corr/MahmudCN14,chen2012you}) and massively multiplayer online role playing games (e.g., EverQuest II; \cite{5284154}), but paradigms like health forums (e.g. DailyStrength) are still mostly unexplored. A number of these social networking services provide their data for nonprofit and research purposes, and there is a huge opportunity to apply data mining and natural language processing in these data to establish successful engagement prediction models for these social networking paradigms.

\section*{Survey Methodology}
We used a handful of keywords related to churn and/or engagement to search for literature appropriate for this survey, for example, \textit{churn}, \textit{continued participation}, \textit{engagement}, \textit{engaged}, \textit{attrition}, etc. These keywords, along with some common keywords related to social networks like \textit{social}, \textit{social media}, \textit{social network}, \textit{newsgroups} and names of popular social networks like Twitter provided a large number of articles in Google Scholar searches. We took the articles from the first page of search results for each of these searches. Of these search results, we discarded all articles that were not about engagement in social media. For example, a large number of the collected articles analyzed social media discussions of churn in telecom. We ended up with only a handful of papers that discussed churn/engagement in social media, and these served as the starting point for our second surveying method. We went through the references of each of these papers to find articles relevant to engagement in social networks. We found articles that were cited over and over by other articles, and ranked them higher in importance for our survey. We again discarded articles that were out of scope, such as research on churn in the service industry. Of the remaining articles, we kept those that were cited by more than one article in our first search. We then conducted the third step in our searching: we went through the articles published by the authors of the already enlisted articles. We selected articles that introduced new techniques, or used combinations of techniques from previous research works in a paradigm not explored before.

\subsubsection*{Domains}
We tried to cover as many social networking domains as possible in this survey. We surveyed research that focused on engagement in traditional social networking websites like Twitter \citep{DBLP:journals/corr/MahmudCN14,chen2012you}, SkyRock \citep{Ngonmang:2012:CPR:2456719.2457022} and Tencent Weibo \citep{liu2017} along with some unconventional ones like UseNet newsgroups \citep{JCC4:JCC433,Arguello:2006:TMF:1124772.1124916}, discussion forums like boards.ie \citep{Karnstedt:2011:EUF:2527031.2527051}, Reddit \citep{hamilton2017loyalty}, Yahoo! Answers \citep{dror2012churn}, health based support forums like DailyStrength \citep{sadeque2015predicting} and HealthBoards \citep{sadeque2016they} etc. We also focused on websites that are not traditionally considered as social networking websites, but has strong inherent social networking properties like massively multiplayer online games, i.e. EverQuest II \citep{5284154} and Top Eleven- Be a Football Manager \citep{Milosevic2017326}, and massive open online courses like Coursera \citep{sinha2014capturing}  .   

\section*{Task Definitions}




The range of tasks that can be performed in the engagement prediction paradigm is quite diverse. The definition of a task depends on three key attributes: types of engagement activities a user is involved in, the definition of future engagement (e.g. length of future engagement, level of engagement in the community etc.), and the influence of peers. A task can try to explore the reception a new user gets from the community and how that affects his or her future participation in the media, or it can try to explore the continuous adaptation process happening to a user during his or her participation in a community and how the lack of adaptability forces a user to disengage, or it can try to look into the diffusion of influence peers have on a user, which may or may not affect the user's future participation in the community. The models and features to be used in the prediction task is almost entirely dependent on how the task is defined.

\paragraph{Engagement as activity performed over time}
The most common form of engagement in this type of research is engagement as activity performed over a certain period of time. Activities may include posts, comments, game sessions, clickstream activities, etc. These studies observe a user's activities over a period of time, and then try to predict the the user's activity level during the following period of time.
In such research, engagement is formally defined as:
\begin{equation*}
engagement_{t,\Delta,\tau}(u) = 
\begin{cases}
   	1 & \text{if } |\{a \in activities(u) : time(a) \in [start(u) + \Delta_o, start(u) + \Delta_o + \Delta_p]\}| > \tau \\
    0 & \text{otherwise}
\end{cases}
\end{equation*}
where $activities(u)$is the set of activities (e.g., posts)  performed by user $u$, $time(a)$ is the time at which activity $a$ occurred, $start(u)$ is the time at which user $u$ first started using the platform, $\Delta_o$ is the \textit{observation period}, the length of time over which the user is observed before trying to make a prediction, $\Delta_p$ is the \textit{prediction period}, the length of time over which engagement is evaluated, and $\tau$ is a threshold indicating how active a user must be to be considered as ``engaged''.

For example, \citet{dror2012churn} considered
$\Delta_o = \text{7 days}$, $\Delta_p = \infty$ and $\tau = 0$:
after observing the first 7 days of a user participating on Yahoo! answers, predict if the user would ever post again to the website.
\citet{sadeque2015predicting} considered
$\Delta_o \in \{\text{1 month, 3 months, \ldots, 24 months}\}$, $\Delta_p = \text{1 year}$ and $\tau = 0$:
after observing a variable number of months (ranging from 1 to 24) of a user on the online health forum DailyStrength, predict if the user would post again to the website within a year.
\citet{Milosevic2017326} considered
$\Delta_o \in \{\text{1 day, 2 days, \ldots}\}$, $\Delta_p = \text{14 days}$ and $\tau = 0$:
after each day a user was active in the online game \textit{Top Eleven - Be a Football Manager}, predict if the user would have any activities in the next 14 days.
\citet{Karnstedt:2010} considered $\Delta_o = \text{26 weeks}$, $\Delta_p = \text{13 weeks}$ and $\tau = T(S) \cdot |\{a \in activities(u) : time(a) \in [start(u) + \Delta_o - \Delta_p, start(u) + \Delta_o]\}|$:
after each activity window ($\Delta_o$) that a user was active in the discussions at boards.ie, predict if the user's activity would be greater than some pre-defined fraction ($0 \leq T(S) \leq 1$) of their activity in the previous activity window ($[start(u) + \Delta_o - \Delta_p, start(u) + \Delta_o]$).
\citet{Danescu-Niculescu-Mizil:2013:NCO:2488388.2488416} measured time in posts rather than calendar units and considered $\Delta_o = \text{20 posts}$, $\Delta_p = \text{200 posts}$ and $\tau = 199$: after observing the first 20 posts of a user, predict if the user posted at least 200 more times.
To make the classification task easier, they discarded users that posted between 30 and 199 times, so that the prediction task only distinguished between users with $<$30 posts and users with $\geq$200.
\citet{JCC4:JCC433}  considered $\Delta_o = \text{time until first post}$, $\Delta_p = \infty$, and $\tau = 0$: given the first post of a user on Usenet newsgroups, predict if the user would ever post again to the website. \citet{Arguello:2006:TMF:1124772.1124916}'s task was defined in the same way.

\paragraph{Engagement as certain amount of activity performed}
Engagement is also commonly defined using the amount of activities performed, regardless of time.
In such research, engagement is formally defined as:
\begin{equation*}
engagement_{f,\tau}(u) = f(u, activities(u))
\end{equation*}
that is, engagement is some function $f$ that maps a set of activities (e.g., posts) performed by user $u$ to a number.

For example, \citet{DBLP:journals/corr/MahmudCN14}'s work on Twitter considered a user to be engaged if their response or retweet rate was above the median rate. This corresponds to defining $f$ as:
\begin{equation*}
f_{X/Y}(u, A) = \frac{|\{a \in A : a \text{ is a } X\}|}{|\{q \in \text{all tweets}: q \text{ is a } Y \text{ by } u\}|} > median(\{f(activities(u') : u' \in U\})
\end{equation*}
where $U$ is the set of all users in the study, $X/Y$ was either response/question or retweet/tweet, and a tweet was considered to be a question if it contained a question mark (`?').

\citet{chen2012you} considered a user to be engaged in the Occupy Wall Street movement based on two different criteria:
how much a user retweeted the official @OccupyWallSt account, and how much they posted tweets with an Occupy Wall Street hashtag (\#OWS).
This corresponds to defining $f$ as one of:
\begin{align*}
f_{RtweerOWS}(u,A) &= |\{a \in A : a \text{ is a retweet of @OccupyWallSt}\}| \\
f_{HashtagOWS}(u,A) &= |\{a \in A : a \text{ contains an \#OWS tag }\}|
\end{align*}



\paragraph{Engagement as loyalty}
\citet{hamilton2017loyalty} tried to predict user \textit{loyalty} to a community, which they define as user preference of a community or a forum over others. This was done by attempting to predict whether a user makes the majority or minority of their posts to a certain forum in a multi-forum site at a specific time.
Formally, engagement in this experiment is defined as:

\begin{equation*}
engagement_{g,t}(u) = \begin{cases}
   	1 & \begin{aligned}
    \text{if } &\dfrac{|\{a \in activities_{t-1}(u) : group(a) = g\}|}{|activities_{t-1}(u)|} \geq 0.5 \text{ and } \\
    &\dfrac{|\{a \in activities_{t}(u) : group(a) =  \in g\}|}{|activities_{t}(u)|} \geq 0.5
    \end{aligned}\\
    0 & \text{otherwise}
\end{cases}
\end{equation*}
Where $activities_t(u)$ are the activities carried out by user $u$ during time window $t$, and $group(a)$ gives the group in which activity $a$ occurred.
A user is thus loyal to group $g$ at time $t$ if the majority of their activities occurred in group $g$ in the preceding time window ($t-1$), and the majority of their activities occurs in group $g$ again at time $t$.



\paragraph{Engagement affected by social influence}
Kawale et al. proposed a churn prediction model that uses social influence among players and their personal engagement in online games \citep{5284154}. They explored the possibility of using diffusion models for predicting future engagement by analyzing influence diffusion among users in the EverQuest massively multiplayer online role-playing game. They observed user interaction for the month of August, 2006, and marked users as churned if they unsubscribed from the game in the following months. Formally,
\begin{equation*}
engagement(u) = \begin{cases}
   	0 , & \text{if } |activities_{2006{\text -}08}(u)| > 1 \text{ and } unsubscribed(u) \in [2006{\text -}08, 2006{\text -}10]\\
    1 , & \text{otherwise}
\end{cases}
\end{equation*}
where $activities_t(u)$ are the activities carried out by user $u$ during time period $t$, and $unsubscribed(u)$ is the time at which user $u$ unsubscribed.



\section*{Models}
A set of analysis and prediction models have been used over the years for future engagement prediction tasks. The learning part of the task is supervised, and a handful of supervised learning techniques (e.g. logistic regression, decision tree etc.) has been used for this. To analyze and understand the importance of attributes in the prediction task, probit models and information gain are used frequently. 

The most common learning technique that has been used in engagement prediction is logistic regression. As the prediction classes are discrete (and intuitively binary, i.e. engaged vs not engaged), using logistic regression makes a lot of sense. Given an input, logistic regression tries to calculate the probability of the input being in either of two classes: a higher probability value means that the input belongs to class 1, whereas a lower value identifies the input as class 0. This is done using a special decision function popularly known as the \textit{Logistic Function}, which is defined as:
$$h_{\theta}(x) = \frac{1}{1 + e^{-\theta^Tx}}$$
Where $x$ is the input vector and $\theta$ is the weight vector for different features of the input learned by the algorithm.

Logistic regression is simple to implement, easy to understand and works really well for data that are not too large and have reasonable dimensionality. A handful of the tasks described above \citep{Danescu-Niculescu-Mizil:2013:NCO:2488388.2488416,chen2012you,dror2012churn,sadeque2015predicting,Milosevic2017326,sadeque2016they} took advantage of these properties, and used logistic regression successfully in their prediction tasks.

Another extremely popular binary supervised learning technique is Support Vector Machine\add{s}, which is a form of regression, but tries to classify samples by establishing an \textit{optimal hyperplane}. A decision hyperplane is optimal if it is optimally distanced from both of the classes, thus creating a \textit{margin} which makes the model less error-prone while classifying new data points. Samples that lie on the margin are called \textit{support vectors}. Although originally intended for linear classification, support vector machines can use \textit{kernels} to perform non-linear classification. A kernel is a function that projects linearly non-separable data points into higher-dimensional feature spaces so that they can be separated using one hyperplane. Introduced by \cite{vapnik} in its current incarnation with soft margins, Support vector machines work really well for binary classification tasks, and have been successfully used by \cite{DBLP:journals/corr/MahmudCN14}, \cite{Ngonmang:2012:CPR:2456719.2457022} and \cite{sinha2014capturing} in their engagement research.

Decision trees are also a really popular supervised learning technique that can be used for discrete classification. Decision tree learning is defined by Tom M. Mitchell as ``a method for approximating discrete-valued target functions, in which the learned function is represented by a decision tree'' \citep{mitchell1997machine}. Decision tree learning usually obtains information gain of an attribute from the attribute set presented with inputs using an impurity measure (e.g. entropy, gini impurity etc.), and builds a tree where attributes with higher information gain are closer to the top. 
Decision trees are easily interpretable, have very fast classification time and can handle both numerical and categorical data, but are computationally complex and may take a long time to train. They are also prone to overfitting. To avoid the problems of overfitting, random forests have been introduced. Random forest learning is essentially an ensemble learning method where random decision trees are created at training time, and classification output of a sample is given by averaging the outputs of the generated trees. This process is much harder to perform, less interpretable and more computationally intensive, but reduces overfitting, and thus is more generalizable. \cite{Karnstedt:2011:EUF:2527031.2527051} used the J48 variation of decision trees, whereas \cite{dror2012churn} and \cite{hamilton2017loyalty} used random forests. \cite{Milosevic2017326} used both decision trees and random forests in their research -- and showed that random forests have a more balanced performance than decision trees.

To analyze importance of attributes of data, \cite{JCC4:JCC433} and \cite{Arguello:2006:TMF:1124772.1124916} used probit analysis. Probit analysis models the marginal change in probability of a binary dependent variable based on the infinitesimal change in continuous independent variables and/or switching values of binary independent variables. Other ways to analyze importance of attributes are information gain \citep{DBLP:journals/corr/MahmudCN14,hamilton2017loyalty}, regression weights \citep{sadeque2015predicting} and pointwise mutual information \citep{sadeque2016they}.

\section*{Feature Types}
Many features have been used to predict continued engagement in social media. The effectiveness of a feature is mostly dictated by the social media that has been researched upon. If the social media is content heavy (e.g. discussion boards), linguistic features can be extremely useful, but as most of these networks have a sparse interpersonal relationship among the users, peer influence or diffusion related features may not be as useful as they are in densely connected social networks like chatrooms or online games. Some features can be ubiquitous in almost all social media like activity frequency or user status, but their effect may not be similar across the platforms. There can be user-level features that contribute to that user's or his peer's future engagement prediction, and there can be community-level features that influence a user's continued participation within that community.

We divide all the features that have been used in the aforementioned tasks into four groups: demographic features, linguistic features, activity features, and interpersonal relationship features.

\section*{Demographic features}
A user's demographic information like age, sex, location etc. can be a predictor of his or her future participation in online social networks.
The intuition behind this is that people from a certain age group tend to use social networking cites more than others, or people from a certain location can be more invested in one particular social network. Unfortunately, no studies have found a concrete correlation between a user's demographics and his or her future engagement in a particular social media. \cite{DBLP:journals/corr/MahmudCN14} tried to observe whether demographic attributes of Twitter users predict their response and retweet rate, and could not find any conclusive evidence. \cite{sadeque2015predicting} used a handful of demographic features (age, sex location) in their participation prediction task, and these features were consistently ranked as low impact across various setups of the experiment. \cite{chen2012you} found some form of correlation between Twitter users' location and their involvement in the \textit{OccupyWallStreet} movement, but this may be the case that the movement was highly centralized in one location (the USA) and people from that location were naturally more invested in it. 

\section*{Linguistic Features}
Linguistic features can contribute to a user's future participation in an online social network. The contents, emotional tone, length of the posts and replies a user has posted in a social network can be good predictors of whether the user is going to leave the forum or not. Also, the responses the user receives from other users can play an important role in the prediction task. These features can be more prevalent in text-heavy social networks like newsgroups or health forums.  

\paragraph{Length of content}
Intuitively, a user who posts more and has more content in a post tends to be identified as an active user, and it is expected that a currently active user will continue participating in the community in the future. This intuition was confirmed by \citet{JCC4:JCC433} - they found that a longer initial post by a user significantly improves prediction. \citet{hamilton2017loyalty} supported this - they found out in most of the subreddits they analyzed, loyal users (users who prefer that subreddit over others) are more verbose than vagrants (users who do not have a particular affinity to a subreddit). \citet{Arguello:2006:TMF:1124772.1124916} contested this finding by presenting that longer sentences in an initial post from a user hurts his or her chance of getting a reply, which in turn reduces the probability of that user posting again in that community. They also found out that longer replies received by a new user can also reduce the post-again probability.

\paragraph{Content complexity}
\label{content}
Complexity of a content plays a crucial role in receiving a response within a community. More complex sentences and vocabulary impose a greater cost on readers, reducing
the chances that the message will be read or responded to \citep{Whittaker:1998:DMI:289444.289500}. Complexity of content can be defined in multiple ways.
\cite{Arguello:2006:TMF:1124772.1124916} used log of message line counts, percentage of long words, and average word count per sentence as  measures of content complexity. \cite{hamilton2017loyalty} introduced esotericity, which is a measure of likeliness of a content to be understood to a select group of people within the community, and is calculated by averaging the inverse document frequency of the noun phrases in the content, that is, 
\begin{equation*}
esotericity_{p} = \frac{1}{|NPs(p)|}\sum_{np \in NPs(p)}{\log\frac{|P|}{|\{p' \in P : np \in p'\}|}}
\end{equation*}
where $NPs(p)$ is noun phrases in post $p$, $P$ is the set of all posts.

The effect of complexity is controversial. \cite{Arguello:2006:TMF:1124772.1124916} showed that more complex initial posts garner less replies, and as the number of replies received has a positive correlation with the post-again probability (see \textit{Activity Features}), getting fewer replies hurts the probability of future engagement. They also showed that getting more complex replies from current user base of the community also discourages a new user to post again. \cite{hamilton2017loyalty}'s finding tells a different story: loyal users of a community prefer posts with more esoteric contents, and a user is more likely to receive responses from the loyal userbase if the content has more esotericity, which in turn influences the user to continue participating.


\paragraph{Psycholinguistic features}
Psycholinguistic features within the posts of a user or replies received by a user can play a crucial role in someone's probability of future engagement. As a participant of a certain ``Gift Economy'' system, a user is likely to contribute more if he is supportive to a community and in turn the community is supportive to him. To analyze psycholinguistic features of a content, the Linguistic Inquiry and Word Count (LIWC) vocabulary is often used. \cite{Arguello:2006:TMF:1124772.1124916} showed that posts that express either positive or negative emotion and/or more cognitive mechanisms are more likely to receive a reply, and if the reply itself has a positive emotional tone, the user is more likely to post again in the community. \cite{DBLP:journals/corr/MahmudCN14} found significant positive correlation between a user's response rate in Twitter and usage of cognition, communication, social process and positive feelings words, and negative correlation with anger and anxiety words. As for retweet rate, they found a positive effect of perception, communication, social process, positive feelings, positive emotions and inclusive words, whereas tentative words have a significant negative effect. These effects were analyzed by \citep{JCC4:JCC433} too, but they found that these effects are highly variable based on the community, and therefore,  not conclusive.

\paragraph{Presence of pronouns}
Presence of certain classes of pronouns can have significant effects on a user's continued engagement in a community. \cite{Arguello:2006:TMF:1124772.1124916} found that presence of sentences containing first person singular pronouns and third person pronouns increased the likelihood of a user posting in that community again. \cite{Danescu-Niculescu-Mizil:2013:NCO:2488388.2488416} also used first person pronouns as a predictor of linguistic adaptability of a user over time. They observed that as a user gets more familiar with an online setting, the use of first person pronoun changes (singular pronouns become plural, e.g. user uses more ``we'' than ``I''), which in turn positively effects users' future engagement in the community. \cite{hamilton2017loyalty} also observed a similar trend, as they found out that in most of the subreddits, loyal users use more second person singular and first person plural pronouns, whereas vagrant users tend to use more first person singular pronouns.

\paragraph{Presence of questions}
As it has been established by \cite{JCC4:JCC433} and \cite{Arguello:2006:TMF:1124772.1124916} that getting a reply in their first post influences new users to continue participating in the social network more, it is intuitive that the first post being a question is a decent way to get replies from existing users. \cite{JCC4:JCC433} showed that if a new user asks a question in their first post, they are more likely to receive a reply, but surprisingly, the reply being actually informative or being another question had no influence on the user's future participation. \cite{Arguello:2006:TMF:1124772.1124916} had similar observations: they found out that if a question asked by a new user is topical to the forum, the user is 6\% more likely to receive a reply. \cite{dror2012churn} had a set of question-related features e.g. question category, length in words and characters, total and average number of questions a user has answered that were deleted, average number of stars
of the questions answered etc. and the only correlation they found between the question features and user churn is that those who participate in a more question-intensive contents are more likely to churn. According to \cite{sadeque2015predicting}, presence of questions in health forum posts indicate future churn of the user, which supports \citeauthor{dror2012churn}'s findings. 


\paragraph{Similarity to language of community}
Similarity between a user's language and a community's language is a fine predictor of the user's future engagement prediction, as higher similarity indicates that the user is adapting well to the norms of the community. \cite{Danescu-Niculescu-Mizil:2013:NCO:2488388.2488416} has presented multiple features that represent some sort of linguistic similarity:
\begin{itemize}
\item \textbf{Language Model Cross-entropy} Average cross-entropy of the post according to the snapshot language model of the month. Cross-entropy according to a bigram language model can be defined as:
\[H(p,SLM_{m(p)}) = -\frac{1}{N}\sum_i \log P_{SLM_{m(p)}}(b_i)\]
where $H(p,SLM_{m(p)})$ is the cross entropy, $p$ is the post whose language is being scrutinized, $m(p)$ is the month when the post was written, $P_{SLM_{m(p)}}$ is the probability of the bigram $b_i$ under the snapshot language model of that month $SLM_{m(p)}$ and $b_1,...,b_N$ are the bigrams of \textit{p}.
\item \textbf{Jaccard self-similarity} Jaccard self-similarity of the current post with past ten posts. This feature provides an insight of linguistic flexibility of the user. It is defined as
\[J(B_c,B_p) = \frac{|B_c \cap B_p|}{|B_c \cup B_p|}\]
where $B_c$ is the set of bigrams of the current post and $B_p$ is the set of bigrams of a previous post.
\item \textbf{Adoption of lexical innovation} is a binary feature: it takes value 1 if the post contains a lexical innovation in the community in the previous three months, and 0 otherwise. Innovation means usage of a bigram that has not been used in the community previously.
\end{itemize}
All these features have positive effects on a user's future engagement in the social network, but there are other linguistic similarity features that have inconclusive effects, for example, \cite{Arguello:2006:TMF:1124772.1124916} found that topical coherence of a post based on bag of words does not have similar effects in all newsgroups they experimented on. 



\paragraph{Other linguistic features}
There are other linguistic features that have been useful in predicting future participation in more than one studies, but are not popular across the domains. For example, \cite{Arguello:2006:TMF:1124772.1124916} and \cite{JCC4:JCC433} both observed that if the first post of a newcomer is testimonial (i.e. post contains user's personal information, or some sort of introduction), it is more likely that the user is going to receive replies. But both of these studies are done on newsgroups, and we do not know how useful it is in other domains.  

\section*{Activity features}\label{activity}
Properties of activities a user performs in social media can be a great predictor of his or her future engagement. Frequency, timing, purpose etc. of activities can contribute to the probability of a user participating in a community. These attributes are relatively easy to observe and analyze, and contain a significant amount of information.

\paragraph{User status}
Status of a user in consideration and also his or her neighbors can be an excellent attribute for predicting continued participation. Status can take multiple forms, e.g. new and old/veteran users (\cite{Arguello:2006:TMF:1124772.1124916,JCC4:JCC433}), loyal and vagrant users (\cite{hamilton2017loyalty}), etc. According to \cite{Arguello:2006:TMF:1124772.1124916}, a newcomer in a social group is less likely to receive a reply from the community on his or her first post, which is an important proponent of the user's future engagement in that network. They also found out that a higher proportion of newcomer-authored replies has a negative effect on future participation. \cite{hamilton2017loyalty} divided users into two cohorts: loyal and vagrant users. Loyal users prefer a group (in this research, a subreddit) over others, whereas vagrants do not have any such identifiable preference. Loyal users tend to be more active in the group they are loyal to, and can effect the future participation of a prospective user by their activities, both positively and negatively.  


\paragraph{Number of activities performed}
One basic attribute that can play an important role in predicting future engagement is the amount of activities performed by a user in the observation period. Activities and observation periods depend on which social networking site the user subscribe to, for example, in online forums or bulletin boards the amount of activity can be just the number of pieces of content a user has posted, whereas in an online gaming platform it can be the number of gaming sessions. Collecting these attributes can be of various degrees of difficulty: collecting the number of posts or gaming sessions can be really straight-forward in online games or support forums \citep{Milosevic2017326,sadeque2015predicting}, whereas churn prediction in massively open online courses required a complex graph architecture of activities \citep{sinha2014capturing}.

In their research on engagement in the \textit{OccupyWallStreet} movement, \cite{chen2012you} used the number of tweets by a follower of the movement as one of the foremost attributes for the prediction task. The number of posts had some significance in \cite{sadeque2015predicting}'s participation prediction task too. \cite{liu2017} also used the number of posts from a user for their churn prediction task. \cite{Milosevic2017326} defined the amount of activity using two features: game session counts and click count within the game. In all of these cases the amount of activity had a positive relationship with future participation, i.e. a higher amount of activity translated into higher probability of future engagement.

\paragraph{Number of replies received}
Number of replies received by a user in his or her activity can be a dominant predictor of that user's future engagement in social media. A higher number of replies from other users in the network can be interpreted as a sign of a welcoming community, and that can play a significant role in a user's future participation. Although the content of the replies, their emotional tones etc. can play crucial roles in the effectiveness of this feature, research has shown that even just the raw amount of replies received can contribute a lot to the prediction task. \cite{Arguello:2006:TMF:1124772.1124916} showed that there is a positive correlation between a higher number of replies received and a higher probability of future engagement in new users. This finding is not consistent with \cite{sadeque2015predicting}'s research, as in their prediction task they found replies received from other users had little to no effect on a user's future participation.   

\paragraph{Temporal characteristics of subscription}
As we have seen previously, the user status, or in other words, subscription duration of a user can be an important attribute towards future engagement. It is crucial to note that the temporal characteristics of this subscription can also be extremely important. Temporal characteristics of subscription can be captured in different ways, e.g. reciprocity, idle time, lurking period, etc. In their research, \cite{sadeque2015predicting} showed that lurking period (time gap between a user's registration and first activity, time gap between a user's last activity and end of an observation period), and average idle time between activities can be useful predictors of a user's future participation in a social media. All three of these features had positive correlation with a user's discontinuation of participation, that is, the higher these numbers were, the more likely that user was going to leave the forum. \cite{Karnstedt:2011:EUF:2527031.2527051} introduced \textit{Reciprocity}, which can be defined as the average time it takes for a post from a user to be replied to, and higher reciprocity (higher average time to garner replies) had significant influence on a user's churning.   

\paragraph{Popularity}
\cite{Karnstedt:2011:EUF:2527031.2527051} defined popularity as "the percentage of the posts a user got replies to". Popularity of a user \textit{i} over time period $[t_1,t_2]$ is defined as:
$$pop_i(t_1,t_2) = \frac{|r(pst_x,pst_y,t_{xy})|}{|Pst_i(t_1,t_2)|}$$
where $Pst_i(t_1,t_2)$ denotes the set of posts user \textit{i} has posted over the time period $[t_1, t_2]$, $r(pst_x,pst_y,t_{xy})$ denotes $pst_y \in Pst(t_1,t_2)$ is a reply of post $pst_x \in Pst_i(t_1,t_2)$ and there was a delay of $t_{xy}$ time units (minutes) between them. The researchers found out that non-churning users enjoy a consistent level of popularity over their user lifecycle in the social media, whereas churners have a more sporadic behavior. 

Another measure for popularity \del{is} introduced by \cite{Karnstedt:2011:EUF:2527031.2527051}, \textit{Initialization}, instead of focusing a user's popularity, focuses on the number of popular threads authored by a user. It is defined as
$$
\operatorname{\mathit{init-pop}}_i(t_1,t_2) =
\frac{|\{thr_l|thr_l \in init_i(t_1,t_2)\wedge |thr_l|>1\}|}{|init_i(t_1,t_2)|}
$$
where $|thr_l|$ is the length of thread \textit{l} in words and $|init_i(t_1,t_2)|$ is the length of the set of the threads initialized by user $v_i$ over time period $[t_1,t_2]$ 
Although intuition suggests that higher initialization should have positive effect on a user\add{'}s not churning from a group, unfortunately, it is not evident from the analysis presented by the researchers.

Popularity can also be measured in other ways, i.e. centrality (defined in the next section), number of followers, number of times a user has been retweeted \citep{chen2012you}\add{,} etc. Whereas higher centrality indicates that a user is more likely to churn, higher number of followers and getting regular retweets has the opposite effect.

\section*{Interpersonal Relationship features}
Interpersonal relationships among the users in a social networking platform is the attribute that largely makes participation prediction in social networks distinct from churn prediction in the telecommunication sector. All social networks have some forms of interpersonal relationships, though some are relatively sparse. Representation of these relationships as graphs can introduce features that can be used to predict future participation in online social networks. Generally, users are represented as nodes in those graphs, whereas edges represent either interpersonal relationships or influence diffusion among these users. These graphs provide researchers with some useful attributes for participation prediction, i.e. degrees of nodes, various types of centrality, graph density, number of strongly connected components and self loops, positive and negative influence diffusion, etc.

\paragraph{Degree}
Degree of a node can be defined as the number of edges incident to that node. The number of incoming edges is called in-degree, and the number of outgoing edges is called the out-degree. In interpersonal relationship graphs, an edge exists between two nodes (users) if there is some sort activity shared between the two users; and the direction of the edge describes the direction of the activity, e.g., if user A replies to a post authored by user B, there is an edge from node A to node B. 

The correlation between the degree of a user and his or her future engagement is always positive. \cite{Karnstedt:2011:EUF:2527031.2527051}, \cite{5284154}, and \cite{Ngonmang:2012:CPR:2456719.2457022} all observed that nodes with higher degrees represent users who are more engaged in the communities.

\paragraph{Centrality}
Centrality can be of two types:
\begin{itemize}
\item \textbf{Closeness Centrality} The importance of a user based on their location in a graph. Let $d_{i,j}$ be the length of the shortest path between vertices $v_i$ and $v_j$. Then average distance between vertex $v_i$ and all vertices is
\[l_i=\frac{1}{|V|}\sum\limits_{j \in V} d_{i,j}\]
Closeness centrality is the inverse of $l_i$
\[C_i = \frac{1}{l_i}\]
\item \textbf{Betweenness Centrality} Measure of a user being a conduit or a broker between communities. Let $\gamma_{x,y}$ be the number of shortest path\add{s} between vertices $v_x$ and $v_y$. Let $\gamma_{x,y,i}$ be the number of those paths where $v_i$ lies on the path, and $v_i \neq v_x$ and $v_i \neq v_y$. Betweenness for $v_i$ is defined as:
\[B_i = \sum\limits_{x,y \in V}\frac{\gamma_{x,y,i}}{\gamma_{x,y}}\]
\end{itemize}
According to \cite{Karnstedt:2011:EUF:2527031.2527051}, both these features play a crucial role in engagement prediction of users in a community. Their analysis showed that users with high centrality measures are more likely to discontinue engagement in the future, and their rationale behind this was that these users are more exposed to the churning of other users in the community as they act as central figures.

\paragraph{Neighborhood Properties}
Neighborhood properties can be the size of the neighborhood, the degree of the neighborhood (neighbors at distance n are n-th degree neighbors), the proportion of churn users within the neighborhood, etc. These features have been used with various degrees of effectiveness. For example, \cite{Ngonmang:2012:CPR:2456719.2457022} observed that neighborhood size has a strong positive correlation with users being engaged in a community. Although not as significant, proportion of inactive users in a user's neighborhood has some effects on his or her future participation. \cite{liu2017} also showed that the higher the proportion of churning users in the neighborhood, the more a user is likely to churn from that community.

\paragraph{Graph Density and Local Clusters}
Graph density can be defined as a ratio of the number of edges and number of nodes in a graph. For an undirected graph, density can be defined as $$D = \frac{2|E|}{|V|(|V|-1)}$$ and for a directed graph, which is the most common type of graph in social media, density can be defined as $$D = \frac{|E|}{|V|(|V|-1)}$$
\cite{hamilton2017loyalty} showed that loyal communities are denser and have less local clustering, indicating that loyal communities are more tight-knit and cohesive. The edge density of the interpersonal graph in loyal communities is significantly higher than that of the non-loyal ones, which indicates that the average user in a loyal community interacts with more users. This interaction is skewed by highly active users though, as these active users communicate with more users on average than the active users in non-loyal communities do. 

\paragraph{Influence Diffusion}
Influence diffusion is defined by the distribution of impact that a user has over his neighbors and how the influence propagates over the graph. \cite{5284154} introduced a Modified Diffusion Model in their paper. In this model, each node on the graph has an influence vector (with direction between nodes) and has a positive (pi) or negative (ni) component (denotes positive and negative influence) based on whether the influencing user is a churner or a nonchurner and represents how much a user is influenced for or against the game. The spread factor $\gamma$ represents the portion of influence that is transferred from a user to the network. The total influence of the graph remains unchanged over propagation of the model, only the positive and negative influence values change. In the paper, Kawale et al. showed the diffusion model's superiority in predicting user-level churn over models that do not consider influence diffusion, as it improved the performance of the prediction model significantly.

\paragraph{Other interpersonal relationship features}
All the aforementioned features are dictated by the communities and social networking sites they were used in by various degrees. Group identity (which group the user has posted in), group size, group-level activity volume, cross-posting characteristics, structure of the community: all of these can influence the effectiveness of certain features, although they are not useful as features by themselves.

\section*{Conclusions}
In this article, we have surveyed the current established works on predicting engagement in social media. It is evident from our survey that, unlike other industries like telecommunication, establishing a universal technique for engagement prediction is difficult because of some unique challenges social media pose due to their variety in structure, user base, communication technique, engagement hierarchy etc. The phenomenon of gift economy also plays a role, while the lack of triggering events makes the task even more complicated. We summed up the most effective techniques used in engagement prediction in social networks based on their task definitions. This study will help future researchers to get a concise view into the current situation of the field and encourage current researchers to explore different approaches and get the best out of them. 

\bibliography{main}

\end{document}